\documentclass[12pt]{article}
\usepackage{amssymb, epsfig}
\newcommand{\newc}{\newcommand}
\newc{\beq}{\begin{equation}}
\newc{\eeq}{\end{equation}}
\newc{\beqa}{\begin{eqnarray}}
\newc{\eeqa}{\end{eqnarray}}
\newc{\bit}{\begin{itemize}}
\newc{\eit}{\end{itemize}}
\newc{\bflr}{\begin{flushright}}
\newc{\eflr}{\end{flushright}}
\newc{\nonr}{\nonumber}
\newc{\hs}{\hskip 2mm}
\newc{\ra}{\rightarrow}
\newc{\Tr}{\mbox{\sl{Tr}}}
\newc{\IM}{\mbox{\sl{Im}}}
\newc{\RE}{\mbox{\sl{Re}}}
\newc{\SU}{$SU(3)_W$ }
\newc{\trip}{$\mathbf{3}$ }
\newc{\sext}{$\bar{\mathbf{6}}$ }
\newc{\tripp}{$\mathbf{3^{\prime}}$ }

\newcommand{\PRD}[3]{{Phys.~Rev.} \textbf{D#1},({#2}) #3}
\newcommand{\PLB}[3]{{Phys.~Lett.} \textbf{B#1}, ({#2}) #3}
\newcommand{\PRL}[3]{{Phys.~Rev.~Lett.} \textbf{#1}, ({#2}) #3}
\newcommand{\NPB}[3]{{Nucl.~Phys.} \textbf{B#1}, ({#2}) #3}

\newc{\PTP}[3]{{Prog. Theo. Phys.} \textbf{#1}, ({#2}) #3}

\begin{document}
\begin{titlepage}
\begin{center}
{\large {\bf  Neutrino Masses in a 5D $SU(3)_W$ TeV  Unification Model }}\\

\vspace{1.2cm}

Chia-Hung V. Chang$^{\ast, \ast \ast}$\footnote{email:
chchang@phy.ntnu.edu.tw}, We-Fu Chang$^{\ast \ast}$\footnote{email: wfchang@triumf.ca}
and
J. N. Ng$^{\ast \ast}$\footnote{email: misery@triumf.ca}

\vspace{0.8cm}
$^{\ast}$ Physics Department, National Taiwan Normal University, Taipei,
Taiwan\\
$^{\ast \ast}$TRIUMF, 4004 Westbrook Mall, Vancouver, BC, Canada V6T 2A3\\[.4cm]
\end{center}

\vspace{1cm}

\begin{abstract}
We study the generation of neutrino masses in the $SU(3)_W$ electroweak unified theory
in $M_4\times S_1/(Z_2\times Z'_2)$
spacetime. By appropriate orbifolding, the bulk symmetry
$SU(3)_W$ is broken into $SU(2)_L \times U(1)_Y$
at one of the fixed points, where the quarks reside. The leptons
form $SU(3)_W$ triplets, localized
at the other symmetric fixed point.
The fermion masses arise from the bulk Higgs sector containing
a triplet and an anti-sextet.
We construct neutrino Majorana masses
via 1-loop quantum corrections
by adding a parity odd bulk triplet scalar.
No right-handed neutrino is needed. The
neutrino mass matrix
is of the inverted hierarchy type.
We show that the model can easily
accommodate the bi-large mixing angle solution favored by the
recent neutrino experiments
without much fine tuning of
parameters. The constraints from $\mu\ra 3e$
transition and neutrinoless double
 $\beta$ decays are discussed.
\end{abstract}

\vspace{2cm}
\vfill
\end{titlepage}
\section{Introduction}
Recently a five dimension (5D) field theory on the orbifold
$S_1/(Z_2\times Z_2^{\prime})$ with bulk $SU(3)_W$
gauge symmetry was proposed to unify the electroweak gauge symmetries of
$SU(2)_L$ and $U(1)_Y$ \cite{HN,lili,DK}. This is a higher dimension version of
an earlier proposal \cite{Wein}. The background geometry  of the
fifth dimension denoted by $y$ is a circle $S_1$ with radius $R$
moded out by two parities and has two fixed points
at $y=0$ and $y=\pi R/2$. At each fixed point a brane is located.
The brane at $y=0$ is $SU(3)_W$ symmetric. On the other hand the one at $\pi R/2$ is not. This is achieved
by orbifold boundary conditions. However, on the second brane  $SU(2)_L \times U(1)_Y$ still holds
and it is
broken by the usual Higgs mechanism.
This unified theory gives a prediction of $\sin^2 \theta_W=0.25$ at the tree level.
The discrepancy with the observed value at $M_Z$ of $\sin^2 \theta_W(M_Z)=0.23$ can be
accounted for by the coupling constant running from the cutoff scale $M^*$ to $M_Z$.
Electroweak unification scale at a few TeV was found to be phenomenologically
viable \cite{HN,DK}.

It is well known \cite{Wein} that the SM doublet and singlet right-handed chiral lepton
can be embedded into a $SU(3)_W$ triplet as given below \cite{notation}
\beq
L_i=\left(\begin{array}{c} e_{i} \\   \nu_{i} \\  e_{i}^c
\end{array}\right)_{L}.
\eeq
On the other hand, the hypercharges of the quarks are too small for similar
embedding into $SU(3)_W$ multiplets. This suggests that the leptons and quarks be located
at different fixed points
in $y$ \cite{DK,Kim}.
Thus, the leptons can be located in the $SU(3)_W$ symmetric brane
at the $Z_2$ fixed point $y=0$ or in the $SU(3)_W$ symmetric bulk. For definiteness we focus
on the brane lepton case. Since the quarks do not form complete multiplets they can
 only be placed  on the $SU(2)\times U(1)$ brane which is
at the $Z_2'$ fixed point $y=\pi R/2$.
This intriguing set up of leptons
points to a violation of the usual additive lepton number conservation scheme. It is more
akin to the forgotten  Konopinski-Mahmoud \cite{KM} assignments. Coupled with recent
progress in orbifold field theories, new possibilities of studying neutrino masses are now opened.
With the lepton number violation, radiatively generated neutrino masses,
similar to the proposal in the Zee model \cite{Zee}, are possible in this scenario.
In extra dimensional models, it is customary to employ one or more
right-handed SM singlet bulk field to generate a small Dirac neutrino masses \cite{fgnu}.
We shall demonstrate here that using only
the minimal number of chiral fermions contained in the SM and appropriate orbifolding, phenomenologically
viable neutrino mass model can be constructed. Since no right-handed neutrinos are introduced,
our construction is fundamentally different from the seesaw mechanism.

\section{ The 5D $SU(3)_W$ Electroweak Model}
We reiterate that the model we study has  only the minimal SM chiral matter fields and bulk
\SU gauge symmetry. However, the Higgs fields are drastically different and
will be discussed in detail later. The extra dimensional space is flat
with orbifold compactification
of $S_1/(Z_2 \times Z_2^{\prime})$. This means that the fifth dimension is the compactified space
$S_1$ of range $[-\pi R,\pi R]$ moded out by two parities $Z_2$ and $Z'_2$.
Under $Z_2$ we have $ y\Leftrightarrow -y$ and $y=0$
is clearly a fixed point. Now relabel the coordinate as $y'=y-\pi R/2 $ and consider
$y'\in[-\pi R/2,\pi R/2]$.
The the second $Z_2^{\prime}$ is the transformation $y'\Leftrightarrow
-y'$. This has fixed points at $y=0,\pi R/2$. The combination of the two $Z_2$
 mappings is equivalent to the
mapping $y\Leftrightarrow y+\pi R$ which is a twist.
These parities can be used to break the
symmetry of the field theory by projecting out even
or odd Kaluza-Klein states under $Z_2$ or $Z_2^{\prime}$ \cite{KA}. This will be explicitly shown later.
Having  defined the geometry we now place the leptons families  at $y=0$ and the
quark families are located $y=\pi R/2$.

Next, we list the bulk Higgs fields we require. First we need a triplet Higgs
$\mathbf{3}$ in order to give lepton masses via Yukawa interactions.
However, the resulting  charged lepton
masses are not realistic and an antisextet
\sext has to
be employed \cite{DK}. For reasons which will be made clear later we also need a second $\mathbf{3}$.
These bulk fields  are
represented by 3-columns $\phi_3,\phi_3^{\prime}$ and a symmetric $3 \times 3$
matrix $\phi_6$ and the bar is dropped for notational
simplicity. The difference between \trip and \tripp is their parity
assignments.
 We use $3 \times 3$ matrices $P,P^{\prime}$ to denote respectively their
parities under $Z_2$ and $Z_2^{\prime}$:
\beqa
\phi_3(y)=P\phi_3(-y), & &  \phi_3(y')=P^{\prime} \phi_3(-y'), \nonr \\
\phi'_3(y)=P\phi'_3(-y), & &  \phi'_3(y')=-P^{\prime} \phi_3(-y'), \nonr \\
\phi_6(y)=P \phi_6(-y) P^{-1}, & &  \phi_6(y')=-P^{\prime} \phi_6(-y') P^{\prime -1}.
\eeqa
$P$ and $P^{\prime}$ are chosen to break bulk \SU symmetry properly,
\beqa
P=\left(\begin{array}{ccc}
 1&0& 0  \\
 0 &1& 0 \\
 0& 0 & 1
\end{array}\right),  \hs
P^{\prime}=\left(\begin{array}{ccc}
 1&0& 0  \\
 0 &1& 0 \\
 0& 0 & -1
\end{array}\right).
\eeqa
With the above assignments, the components of the Higgs multiplets and their parities are
\beqa
\phi_3=\left(\begin{array}{c}  \phi_3^-(++)\\\phi_3^0(++)\\ h_3^+(+-)
\end{array}\right),\hs
\phi'_3=\left(\begin{array}{c}  \phi_3^{'-}(+-)\\\phi_3^{'0}(+-)\\ h_3^{'+}(++)
\end{array}\right),
\label{P3}
\eeqa
and
\beqa
\label{P6}
\phi_6=\left(\begin{array}{ccc}
 \phi^{++}_{\{11\}}(+-) &\phi^{+}_{\{12\}}(+-)& \phi^{0}_{\{13\}}(++)  \\
 \phi^{+}_{\{12\}}(+-) &\phi^{0}_{\{22\}}(+-)& \phi^{-}_{\{23\}}(++) \\
 \phi^{0}_{\{31\}}(++) &\phi^{-}_{\{32\}}(++)& \phi^{--}_{\{33\}}(+-)
\end{array}\right).
\eeqa
They are not used to break the \SU symmetry spontaneously. Instead they play the role of generating
fermion masses.
 The parities given above is engineered to give a reasonable
mass pattern for the leptons in the lowest order.
Under the assignment, only the parity positive $\phi_3^3$ and $\phi^{0}_{\{13\}}$
could develop nonzero vacuum expectation value (VEV) and generate the charged lepton masses.
This is the central ingredient
in orbifold treatments of the flavor problem.
To see this more clearly we need to construct
the 5D Lagrangian density which is
invariant under \SU and the orbifold symmetry. It is given by
\beqa
{\cal L}_5 &=& -\frac14 G^{(a)}_{MN}G^{(a)MN}
 +\Tr[(D_M\phi_6)^\dag(D^M\phi_6)]+(D_M\phi_3)^\dag(D^M\phi_3)
 +(D_M\phi'_3)^\dag(D^M\phi'_3)\nonr\\
 && +\delta(y) \left[\epsilon_{abc}\frac{ f_{ij}^3}{\sqrt{M^*}} \overline{(L^a_i)^c} L^b_j \phi_3^c
 + \epsilon_{abc}\frac{f_{ij}^{'3}}{\sqrt{M^*}}  \overline{(L^a_i)^c} L^b_j \phi_3^{'c}
 + \frac{f^6_{ij}}{\sqrt{M^*}}\overline{(L^a_i)^c} L^b_j \phi_6^{\{ab\}}
 + \bar{L}\gamma^\mu D_\mu L\right]\nonr\\
 && -V_0(\phi_6,\phi_3,\phi'_3) -
 \frac{m}{\sqrt{M^*}}\phi_3^T \phi_6 \phi'_3 + H.c.
 +\mbox{quark sector}.
\label{5DL}
\eeqa
The notations are self explanatory. The cutoff scale $M^*$ is
 introduced to make the coupling constants
 dimensionless. In the literature, the strong coupling requirement
 is usually employed to fixed the ratio $M^* R \approx 100$ (see, for example, \cite{HN}).
 The quark sector is not relevant now and will be left
out. The complicated scalar potential is gauge invariant and orbifold symmetric and will
not be specified.

The 5D covariant derivatives are
\beqa
D_M\phi_3= (\partial_M+ i g' A^{a}_M T^{a})\phi_3,\\
D_M\phi_6= \partial_M \phi_6+ i g' [ A^{a}_M T^{a}\phi_6+
\phi_6 (A^{a}_M T^{a})^T ]
\eeqa
with generator $T^{a}=\frac12 \lambda^{a}$.
The gauge matrix ${\cal A}_M \equiv A^{a}_M T^{a}$ is:
\beq
{\cal A}=\frac12
\left(
\begin{array}{ccc}
 $$A^3+{1\over \sqrt3} A^8$$ & $$\sqrt{2} T^+$$ & $$\sqrt{2} U^+$$ \\ %
 $$\sqrt{2} T^-$$  & $$-A^3+{1\over \sqrt{3}} A^8$$ & $$\sqrt{2} V^+$$  \\
 $$\sqrt{2} U^-$$  & $$\sqrt{2} V^-$$ & $$ -{ 2\over \sqrt{3}} A^8 $$
\end{array}\right),
\eeq
where
\[
T^\pm= {A^2\mp iA^3 \over \sqrt2},\hs
U^\pm= {A^4\mp iA^5 \over \sqrt2},\hs
V^\pm= {A^6\mp iA^7 \over \sqrt2}.
\]
The parities of gauge field are assigned as:
\beqa
&&{\cal A}_\mu(y)=P {\cal A}_\mu(-y) P^{-1},
\hs {\cal A}_\mu(y')=P^{\prime} {\cal A}_\mu(-y') P^{\prime -1},
\hs \mu=0,1,2,3\nonr\\
&&{\cal A}_5(y)=-P {\cal A}_5(-y) P^{-1},
\hs {\cal A}_5(y')=-P^{\prime} {\cal A}_5(-y') P^{\prime -1}.
\eeqa
Explicitly, $U,V$ are assigned $(+,-)$ parities and their Kaluza-Klein decompositions are
\beq
{2\over \sqrt{\pi R}} \sum_{n=0} A^{2n+1}(x) \cos{(2n+1)y\over R}.
\eeq
 It can be seen that their wavefunctions vanish at the fixed point $y=(\pi R/2)$
where quarks live on.
They have no zero modes and their masses are naturally heavy and of order $1/R$.
The remaining entities $A^3, A^8, T^\pm$ are endowed with even parities $(+,+)$ and
have zero modes and they decompose as
\beq
{2\over \sqrt{\pi R}}\left[A_0/ \sqrt2 +\sum_{n=1} A^{2n}(x) \cos{2n y\over
R}\right].
\eeq
The zero modes are identified as the SM gauge bosons.
The bulk Lagrangian still respect a restricted \SU gauge symmetry with the
gauge transformation parameters obeying the same boundary condition as the gauge fields.
Hence, at the fixed point $y=(\pi R/2)$ the gauge symmetry \SU
is reduced to $SU(2)_L \times U(1)_Y$, allowing for the existence of quarks.
The 4D effective Lagrangian can be
obtained from Eq.\ (\ref{5DL}) by integrating out $y$. In particular, we have
the following gauge interactions
\beqa
{\cal L}_g= {i g'\over \sqrt{2\pi M^* R}} \left[\overline{e_L}\gamma^\mu(A^3_\mu+\frac{1}{\sqrt{3}}A^8_\mu) e_L
- \overline{\nu_L}\gamma^\mu(A^3_\mu - \frac{1}{\sqrt{3}}A^8_\mu)
\nu_L\right.\nonr\\
\left.-\frac{2}{\sqrt3} \overline{e_R}\gamma^\mu e_R A^8_\mu
+\sqrt{2} \overline{e_L}\gamma^\mu \nu_L T^-_\mu+ H.c.\right],
\eeqa
and for the KK modes, there is a $\sqrt2$ enhancement factor. The  5D gauge coupling $g'$ is now related to
the $SU(2)$ gauge coupling $g$ at low energy as ${g' \over \sqrt{\pi M^* R}}=\frac{g}{\sqrt{2}}.$
It is important to note that we also have the following interaction
\beq
{\cal L}_{UV}= {i g'\over \sqrt{\pi M^* R}}\left[\sqrt{2} \overline{e_L}\gamma^\mu e_R^c U_{n\mu}^{-2}
+ \sqrt{2} \overline{\nu_L}\gamma^\mu e_R^c V_{n\mu}^{-1} + H.c.\right].
\label{UVgauge}
\eeq
which can induce spectacular lepton number violating effects.  The superscripts on $U$
and $V$ denotes their respective charges.

 It can be seen from Eq.\ (\ref{P3}) and Eq.\ (\ref{P6}) that
only the $SU(2)_L$ doublets in \trip and \sext and the $SU(2)_L$ singlet in \tripp
have zero modes.
Parities and charges allow for the bulk fields $\mathbf{3}$ and
$\bar{\mathbf{6}}$ to develop vacuum expectation values but not the $\mathbf{3^{\prime}}$.
Hence, we have
\beqa
\langle\phi_3\rangle=\frac{v_3^{3/2}}{\sqrt2}
\left(\begin{array}{c} 0\\ 1\\0 \end{array}\right),\hs
\langle\phi_6\rangle=\frac{v_6^{3/2}}{\sqrt2}
\left(\begin{array}{ccc}0& 0&1\\0&0&0\\1&0&0 \end{array}\right).
\label{vev}
\eeqa
A linear combination of the $SU(2)_L$ doublet in the \trip and the \sext then
breaks the SM gauge symmetry. Then the tree level  $W$ boson mass is given by
\beq
M_W^2= {g^{'2} \over 2M^*}\left(v_3^{3}+2v_6^{3}\right)={g^2\pi R(v_3^{3}+2v_6^{3})\over 4}.
\eeq

The charged lepton mass matrix
in the basis of $(e, \mu, \tau)$ can be expressed as:
\beqa
&&\frac{v_3^{3/2}}{\sqrt{2M^*}}(\overline{e_R}, \overline{\mu_R}, \overline{\tau_R})
\left( \begin{array}{ccc}
  0 & f^3_{12} & f^3_{13} \\  - f^3_{12}& 0 & f^3_{23} \\  - f^3_{13}& -f^3_{23} & 0
\end{array}\right)
\left(\begin{array}{c}  e_L\\ \mu_L \\ \tau_L
\end{array}\right)\nonr\\
&+&\frac{v_6^{3/2}}{\sqrt{2M^*}}(\overline{e_R}, \overline{\mu_R}, \overline{\tau_R})
\left( \begin{array}{ccc}
  f^6_{11} & f^6_{12} & f^6_{13} \\
  f^6_{12}& f^6_{22} & f^6_{23} \\
  f^6_{13}& f^6_{23} & f^6_{33}
\end{array}\right)
\left(\begin{array}{c}  e_L\\ \mu_L \\ \tau_L
\end{array}\right)+ H.c.,
\label{lepmass}
\eeqa
 Eq.\ (\ref{lepmass}) shows clearly that \trip alone gives the wrong mass pattern. The correct masses
 will require a detail numerical study of the Yukawa couplings $f^6_{ij}$ and $f^3_{ij}$
which is beyond our scope now. It suffices to note that a correct hierarchy for the charged lepton
masses requires  $(f_3 v_3 / f_6 v_6) \lesssim 0.1$. Thus, to a good approximation, the
charged lepton mass matrix is dominated by the \sext:
\beq
{\cal M}_{ij} \sim f^6_{ij}{M_W\over g'\sqrt2}={f_{ij}^6\over \sqrt{\pi R M^*}}{M_W\over g},\hs
v_6^{3/2}\sqrt{\pi R}\sim v_6^{3/2}\sqrt{\pi R} \sim v_0=250 \mbox{GeV}.
\label{chargemass}
\eeq
Next, we turn our attention to neutrino masses.

\section{5D Model of Neutrino Masses}
\begin{figure}
\epsfxsize=360pt
\centerline{\epsfbox{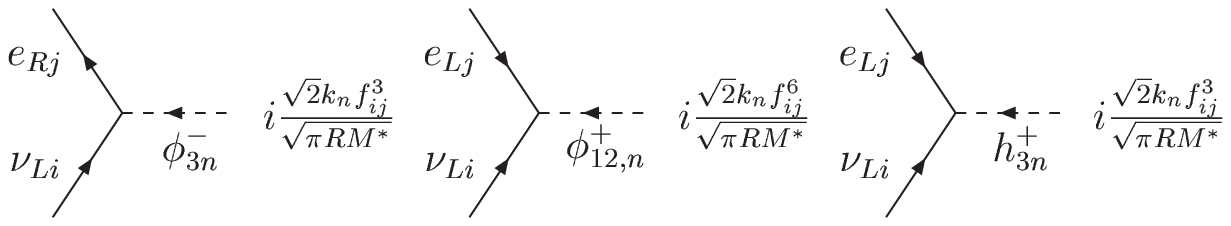}}
\caption{The Feynman rules for the lepton Higgs  couplings,
$i,j$ are the flavor indices, $f^3_{ij}=-f_{ji}^3$, $f^6_{ij}=f_{ji}^6$,
$n$ is the KK number and $k_n=1$ for $n=0$ and $k_n=\sqrt{2}$ for $n\neq0$.
For $\mathbf{3}'$, simply substitute the $f^3_{ij}$ by $f^{'3}_{ij}$.}
\label{fig:vertex}
\end{figure}

The parities given in Eq.\ (\ref{P3}) and Eq.\ (\ref{P6}) disallow  $\phi^0_{\{22\}}$ from developing a VEV
and naturally forbid tree level neutrino masses.
However, the model naturally generates neutrino masses via 1-loop quantum effects. The 4D effective
interaction of the brane neutrinos and the bulk Higgs fields are given by the Yukawa terms of
Eq.(\ref{5DL}). It is
\beq
{\cal L}_{4Y}=\sum_{n}{2 (\sqrt{2})^{-\delta_{n,0}}\over \sqrt{\pi R M^*}}
\left[\epsilon_{abc} f_{ij}^3 \overline{(L^a_i)^c} L^b_j \phi_{3n}^c
 + \epsilon_{abc}f_{ij}^{'3}  \overline{(L^a_i)^c} L^b_j \phi_{3n}^{'c}
 + f^6_{ij}\overline{(L^a_i)^c} L^b_j \phi_{6n}^{\{ab\}}
+H.c.\right],
\eeq
where $n$ is the KK-number and there is a $\sqrt2$ factor enhancement for
nonzero modes. Immediately, we notice that the extra
space volume dilution factor $\sqrt{\pi RM^*}$ naturally show up to suppress
the Yukawa couplings. The Feynman rules for these vertices are
depicted in Fig.\ref{fig:vertex}.

\begin{figure}
\epsfxsize=360pt
\centerline{\epsfbox{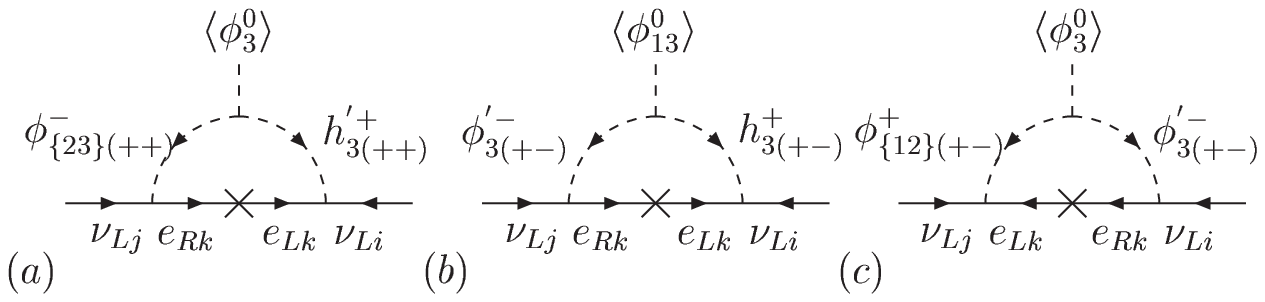}}

\caption{The 1-loop neutrino mass through
$\phi_3^T\phi_6\phi'_3$ coupling.}\label{fig:Zeeloop2}
\end{figure}

The next important ingredient is the \trip \tripp \sext term in Eq.(\ref{5DL}).
 It is this term that violates the usual additive lepton number conservation
and makes the 1-loop Majorana mass possible.
The effective
Higgs mixing is derived to be
\[
-{\sqrt2  m\over \sqrt{\pi R M^*}} \phi^T_{3p} \phi_{6q}\phi'_{3r},
\]
where indices $p,q$ and $r$ stand for the KK numbers which satisfy $|p\pm q\pm r|=0$. When one of the
fields develops a VEV it is replaced by $m(v_b^3/2M^*)^{1/2}$ . These interactions
induce three possible 1-loop diagrams for
generating neutrino Majorana masses, see Fig.\ref{fig:Zeeloop2}. The neutrino mass matrix
is necessarily Majorana since only left-handed neutrinos exist in this model.

We first observe that the dominant contribution comes from Fig.\ref{fig:Zeeloop2}(a) which
is mediated by two Higgs zero modes.
This by itself gives a neutrino mass matrix which is Zee model like \cite{Zee} in
its structure assuming the charged lepton
 mass matrix is diagonal. We can do this without loss of generality and including
charged lepton rotations will only complicate the formulae without adding new insights
to the physics.

 Without further ado,
the elements of the neutrino mass matrix calculated from this diagram is
\beq
({\cal M})^{(a)}_{ij}={1\over 16\pi^2}{2 m(v_3)^{3/2} \over (\pi R M^*)\sqrt{2M^*}}
\sum_k { m_k  f_{ik}^{'3}f_{jk}^6\over M_1^2-M_2^2  }\ln\frac{M_2^2}{M_1^2}
\label{Ma}
\eeq
where $m_k$ is the mass of charged lepton-$k$ and
$M_1,M_2$ are the masses of $h_3^{'+}$ and $\phi_{\{23\}}^-$. Substituting the $f^6_{ij}$
for lepton masses we get to the first order a neutrino mass matrix that is Zee model like:
\beq
{\cal M}_\nu \sim { g\over 16\pi^2 }\frac{v_0}{M_W }
{\sqrt2 m \over(\pi RM^*) }{\ln\frac{M_2^2}{M_1^2}\over (M_1^2\!-\!M_2^2)}
\left( \begin{array}{ccc}
  0 & f^{'3}_{12}(m_\mu^2\!-\!m_e^2) &  f^{'3}_{13}(m_\tau^2\!-\!m_e^2)\\
f^{'3}_{12}(m_\mu^2\!-\!m_e^2)  & 0 &  f^{'3}_{23}(m_\tau^2\!-\!m_\mu^2)\\
f^{'3}_{13}(m_\tau^2\!-\!m_e^2) &f^{'3}_{23}(m_\tau^2\!-\!m_\mu^2) &0
\end{array}\right).
\eeq
If the Yukawa couplings observe the following  hierarchy
\[
f^{'3}_{12}:f^{'3}_{13}:f^{'3}_{23}\sim 1:\epsilon:\epsilon^2,\hs
\epsilon=m_\mu^2/m_\tau^2
\]
then it leads to  bi-maximal mixing of neutrinos \cite{Zeep} which is close but do not
explain the recent data \cite{SK}. It can serve as the leading order approximation
to a more realistic mass matrix.
With the volume dilution factor,
it is very natural to have a small neutrino mass.
As an example, the following parameter set,
\beq
(\pi R M^*) \sim 100.0,\hs
M_2 \sim 300 \mbox{GeV},\hs
M_1 \sim 900 \mbox{GeV},\hs
m\sim 250 \mbox{GeV},\hs
f_{12}^{'3}=- 0.026.
\label{eq:overall}
\eeq
gives the mass scale for neutrinos $m_{\nu 1}\sim 0.06$ eV.
We have normalized the Yukawa coupling with $M_W$, so that $|f^{'3}_{12}|=0.026$
is not unnatural compared with $f^6_{\tau\tau}\sim 0.04$. Also, it
has nothing to do with charged lepton masses and is basically a
free parameter.

The model has a natural perturbation to the Zee mass pattern. They come from the
diagrams of Fig.\ref{fig:Zeeloop2}(b,c). Because they involve KK-Higgs
running in the loop, these diagrams are expected to be  smaller
compared to Fig.\ref{fig:Zeeloop2}(a). Diagram-(c) gives the same
structure as diagram-(a) but suppressed by the KK masses,
 ${\cal M}^{(c)}\sim 2M^2R^2 {\cal M}^{(a)}$, where $M$ represents
 the mass of zero mode Higgs boson in diagram-(a).
Diagram-(b), on the other hand, exhibits different structure and hence can give the perturbation
needed to
account for the data.
The contribution from
diagram-(b) can be calculated from the previous calculation Eq.\ (\ref{Ma}) by replacing $f_6$ with $f_3$,
substituting the zero mode masses by $n-$th KK masses, and inserting the factor $(\sqrt2)^2$
for the normalization of KK modes:
\beqa
{\cal M}^{(b)}
\sim \frac{1}{4 \pi^2 }{m v_0 R^2 \over (\pi RM^*)^{3/2} }\times\nonr\\
\left(\begin{array}{ccc}
 2(f^3_{12}f^{'3}_{21}m_\mu+ f^{'3}_{13}f^3_{31}m_\tau) &
 (f^3_{13}f^{'3}_{32}+f^{'3}_{13}f^{3}_{32} )m_\tau &
 (f^3_{12}f^{'3}_{23}+f^{'3}_{12}f^{3}_{23} )m_\mu\\
 (f^3_{13}f^{'3}_{32}+f^{'3}_{13}f^{3}_{32} )m_\tau &
 2(f^3_{21}f^{'3}_{12}m_e+ f^{'3}_{23}f^3_{32}m_\tau) &
 (f^3_{21}f^{'3}_{13}+f^{'3}_{21}f^{3}_{13} )m_e\\
 (f^3_{12}f^{'3}_{23}+f^{'3}_{12}f^{3}_{23} )m_\mu &
  (f^3_{21}f^{'3}_{13}+f^{'3}_{21}f^{3}_{13} )m_e &
 2(f^3_{31}f^{'3}_{13}m_e+ f^{'3}_{32}f^3_{23}m_\mu)
\end{array}\right).
\eeqa
For simplicity, we only include the contribution from $n=1$ KK states.
Assuming that $f^6_{ij}$ is nearly diagonal
the six couplings $f^3,f^{'3}$ can be adjusted to
fit the neutrino oscillation data.
 As a  first step, we find that to fit all the data, including  the recent KamLAND result \cite{KamL},
the couplings $f^3,f^{'3}$ have a pattern.
We propose the following  parameter
set:
\beqa
\{f_{12}^{'3},f_{13}^{'3},f_{23}^{'3}\}&=& 0.026\times \{-1, 0.75 \epsilon, 0.5\epsilon\},\nonr\\
\{f_{12}^{3},f_{13}^{3},f_{23}^{3}\}&=& 0.090 \times \{-0.1, -0.1, 1.0\}.
\label{tripy}
\eeqa
Here we take $1/R=2$ TeV \cite{HN} and
keep the other parameters the same as in Eq.(\ref{eq:overall}).
It produces the following neutrino mass matrix
\beq
{\cal M}_\nu \sim
\left(\begin{array}{ccc}
  0.420 & 1.0 & 0.922 \\ 1.0 & 0.097 & -0.464 \\ 0.922& -0.464 &
  0.006 \end{array}\right)\times 0.0441 \,(\mbox{eV}).
\label{nusample}
\eeq
This translates into $\theta_{12}=36.6^\circ$,  $\theta_{23}=42.4^\circ$,
$\sin\theta_{13}=0.064$, for the neutrino mixing angles in standard notation,
 and $\triangle M_\odot=7.3\times 10^{-5}(\mbox{eV})^2$
and $\triangle M_{atm}=3.4\times 10^{-3}(\mbox{eV})^2$, for mass square differences.
This pattern is close to the phenomenologically studied inverted mass hierarchy with large mixing angle
solution to the solar neutrino problem given in \cite{HDN}.

It is interesting that the model we constructed is naturally of the inverted
hierarchy kind and a mass at $.06$ eV without excessive fine tuning of parameters.
It is interesting to note that the model cannot  accommodate the normal hierarchy even with fine tuning.

\section{Rare $\mu$ Decays}
The model has lepton number violating  gauge interactions (see Eq.(\ref{UVgauge}) ) as well as Higgs
interactions. The latter arise because the charged leptons get their masses from
the VEV's of both the \trip and the \sext as given in Eq.(\ref{lepmass}). Diagonalization of
${\cal M}_{lept}$ in general does not separately diagonalize the matrices $f^3$ or $f^6$. If we denote the bi-unitary
rotations that diagonalize ${\cal M}_{lept}$ by $U_{L/R}$, the interaction of
neutral Higgs boson with the charged lepton mass eigenstates is
\beq
\left(\frac{\sqrt2}{\sqrt{\pi RM^*}} \right)\bar{l'}_R \left[
\left(U_R^\dag \{f^3_{ij}\} U_L\right)\phi_3^0
+\left(U_R^\dag \{f^6_{ij}\} U_L\right)\phi^0_{\{13\}}
\right] l'_L + H.c..
\eeq
In our scenario, we assume that $v_3 \sim v_6$ and the charged lepton mass hierarchy is due to
$f^3 \ll f^6$ which admittedly is  fine tuning. The $U$ rotations approximately diagonalize the $f^6$ matrix.
Hence, the only flavor changing neutral current comes from the \trip and will be suppressed by
$f^3/f^6$.

\begin{figure}
\epsfxsize=240pt
\centerline{\epsfbox{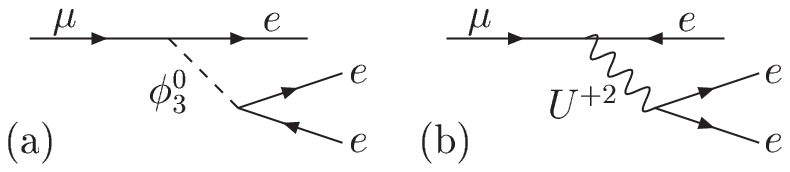}}

\caption{The tree level diagram for $\mu\ra 3e$ induced by (a) the off diagonal couplings
of $\phi_3$ and (b)the  KK $U^{+2}$ gauge boson.}
\label{fig:mu3e}
\end{figure}

Consider the rare decay $\mu \ra 3e$. It can proceed through neutral Higgs exchange or the doubly
charged KK gauge boson $U^{\pm 2}$ as seen in Fig.(\ref{fig:mu3e}).

We estimate the contribution due to $\phi_3^0$ to be given by
\beq
{Br(\mu\ra 3e)\over Br(\mu\ra e \bar{\nu_e} \nu_\mu)}
\sim
6.3\times10^{-17}|\xi_{\mu e}|^2 \left(\frac{ f^3 M_W}{f^6 M_H}\right)^4
\eeq
where we have used  $(\pi R M^*)=100.0$.
Without taking account of the suppression of mixing, $\xi_{\mu e}$, and smallness
of Yukawa couplings this estimate  is already way below the
current experimental bound, $Br(\mu\ra3e)<1 \times 10^{-12}$\cite{PDG}.

 The contribution of $U^{\pm 2}$
can be gleamed from Eq.(\ref{UVgauge}). We note that in the limit of  $f^3=0$ the  mass matrix of
 charged leptons is symmetric which
totally comes from the VEV of Higgs sextet. In that limit, $U_L= U_R^*$ and
there is no FCNC medicated by doubly charged $U^{\pm 2}$ boson, namely
\[
 (U^\dag_LU_R^*)_{ij} = \delta_{ij}.
\]
 When the Yukawa coupling of Higgs triplet is turned on, we expect the
off-diagonal couplings are proportional to $(f^3/f^6)$. Thus, we can give an
 order of magnitude estimate for the branching ratio
\beq
{ Br(\mu\ra 3e) \over Br(\mu\ra e \bar{\nu_e} \nu_\mu) }\sim
{|\frac{g^2}{M_U^2}(U^\dag_LU_R^*)_{ee}(U^\dag_LU_R^*)_{\mu e}|^2 \over |\frac{g^2}{M_W^2}|^2}
\sim ( R M_W )^4 \left(\frac{f^3}{f^6}\right)^2 < 10^{-12}.
\label{mu3eg}
\eeq
Thus this decay can be suppressed by either the compactification scale and/or the ratio
$f_3/f_6$. The compactification scale is usually determined by requiring
the coupling constant running between $M^*$ and $M_Z$ gives the correct prediction
of $\sin^2\theta_W(M_Z)$. For non-supersymmetric version, $M_c$ is predicted to be a few TeV \cite{HN}.
There is not much room to maneuver. To stay below the experimental bound will require
$f_3/f_6 \lesssim 6\times  10^{-4}$ or a special Yukawa pattern which leads
to small $\mu-e$ mixing after mass diagonalization.
 However, for the supersymmetric scenario, $M_c$ could be as large as 100 TeV \cite{Kim},
though the exact number depends on
the detail of sparticle spectrum.

\section{Neutrinoless double beta decays}
\begin{figure}
\epsfxsize=360pt
\centerline{\epsfbox{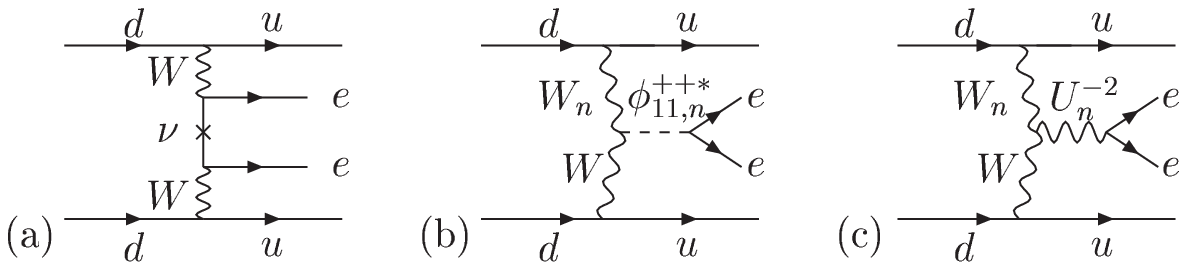}}
\caption{The tree level diagram for $0\nu\beta\beta$ decays
through (a) neutrino Majorana mass and the mediation of KK modes
of (b)$\phi_{\{11\}}^{++}$ Higgs boson and (c) $U^{-2}$ gauge boson.}
\label{fig:0nubb}
\end{figure}

Neutrinoless double beta decay is an important tool in the study of neutrino masses.
A recent analysis taken into account all the recent neutrino data is given in \cite{Petcov}. For
our model there are three
possible sources that can lead to the decay:
\begin{enumerate}
\item
The first entry in the Majorana neutrino mass matrix
\item
The triple coupling of $W^- W^- \phi_{\{11\}}^{++}$
\item
The triple coupling of $W^- W^- U^{+2}$
\end{enumerate}
These are depicted in Fig. \ref{fig:0nubb}.

 Now we can argue that only the one through neutrino mass
is important. The six fermions operator responsible for the process is
 $(\bar{u}d)(\bar{u}d)\bar{e^c}e$ and the coefficient associated with
can be estimated.
From the neutrino mass term we have
\beq
G_{(a)}\sim {g^4 \over M_W^4}{m_\nu \over \langle p^2\rangle}
\eeq
where $p$ is the momentum transfer in this process,
and similarly we have
\beqa
G_{(b)}&\sim& \frac{g^2}{M_W^2}(R^2)\left( {g M_W \over 1/R^2}\right)
\left(g \frac{m_e}{M_W}\right)= g^4 {m_e R^4 \over M_W^2},\\
G_{(c)}&\sim& \frac{g^2}{M_W^2}(R^2)\left( {g \langle p \rangle \over 1/R^2}\right)
(g )= g^4 {\langle p\rangle R^4 \over M_W^2}
\eeqa
for diagram (b) and (c). So their relative size compared to diagram(a) are
\beq
{G_{(b)}\over G_{(a)}}\sim (\frac{m_e}{m_\nu}) M_W^2 \langle p^2\rangle R^4, \hs
{G_{(c)}\over G_{(a)}}\sim (\frac{\langle p\rangle }{m_\nu}) M_W^2 \langle p^2\rangle R^4.
\eeq
Where $\langle p^2\rangle$ is around the order of $10^3-10^4
(\mbox{MeV})^2$, $m_\nu \sim 0.06$eV, and taking $1/R\sim 2$TeV as
example, we have the ratios:
\beq
{G_{(b)}\over G_{(a)}}\sim 6 \times 10^{-6}, \hs
{G_{(c)}\over G_{(a)}}\sim 2.5\times 10^{-3}.
\eeq
Thus, the neutrinoless double beta decays rate is
mainly controlled by the $(11)$ component of neutrino mass matrix.
Due to the charged lepton mass suppression, the contribution
from the physical charged Higgs bosons can be ignored. The example
we have given in Eq.(\ref{nusample}) has  $m_{11}\sim 2\times 10^{-2}$eV. This is well
within the current experimental limit of \cite{2beta}
\[ |\langle m_\nu \rangle| =\left(0.39 {+0.45 \atop -0.34 }\right) \mbox{eV}. \]
Interestingly this may be within reach of the next generation of these experiments.

\section{Conclusions}

We have investigated neutrino masses in the framework of brane world scenario with
\SU bulk symmetry and orbifold symmetry breaking. Since the leptons are localized on
a brane and form a complete triplet, the masses of the charged leptons call for the use of
\trip and \sext Higgs boson when Yukawa interactions are employed. However, the necessary
fine tuning of Yukawa couplings is not explained by the model and is no better understood
than in the SM. One possible way to rationalize  the charged
lepton mass hierarchy is to incorporate the split fermion\cite{SplitF}
with this model which is beyond our scope now.
To generate neutrino masses via orbifold mechanism we have to introduce another triplet, \tripp,
which is odd under the $Z_2^{\prime}$ parity. This is a scalar field and do not develop
VEV. In this way the neutrino masses arise from a 1-loop process. The overall scale is in  the
$0.01$ eV range. Here the suppression comes from the loop integration, the bulk
volume dilution and the smallness of Yukawa couplings, required since the charged lepton
masses are small compared to the weak scale. No other fine tuning is required. The mass pattern
we obtained is of the inverted mass hierarchy type. At the 1-loop level the dominant structure
 of the neutrino mass matrix is Zee model like. However, unlike the Zee model the diagonal
elements are not vanishing but only subleading. This is due to the fact that
they arise from virtual KK particle
exchanges. They have the effect of
inducing  a perturbation to the dominant structure and gives the necessary perturbation
in order to account for the data \cite{HDN}.  The unknown parameters here are the Yukawa couplings
 of the \trip and \tripp.
Since they are not involved with charged lepton mass generation,
which is assumed to be done mainly by a diagonal Yukawa couplings of the \sext,
these parameters are not constrained.
We found that
all the current neutrino oscillation data can be easily accommodated.
One possible pattern was given in Eq.(\ref{tripy}). We digress here for the need to introduce \tripp.
Actually \trip could play the role of \tripp in all the loop diagrams.
However, the triple scalar term \trip \sext \trip is odd under $Z_2 \times Z_2^{\prime}$. Even if we
allow for such terms, we cannot generate a correct MNS neutrino mixing matrix \cite{MNS}.

Returning to our model, it  predicts a small
neutrinoless double decay rate which can be seen from the neutrino mass matrix. This comes from  an  overall
scale discussed above and a small $(11)$ element of the neutrino mass matrix. Our solution also
 indicated a small
value for  $U_{e3}=0.064$ of the MNS matrix \cite{MNS}
and a small $m_{\nu e}\sim 0.06$eV which is well below the upper bond $m_{\nu
e}<2.2$eV from tritium beta decay\cite{tritium}.
In the course of this
study we find no solution that accommodates the data with a normal mass hierarchy. Indeed this
is a generic feature of this scenario.
Although  we have not achieved an understanding of
charged lepton masses our modest attempt in building a model for the masses of brane neutrinos is
nevertheless interesting
since no right-handed neutrinos are introduced. It is an alternative to the current models
of neutrino masses either in four or higher dimensions.

Viewed in the usual additive lepton number picture, this model has many sources that can lead
to lepton flavor violating neutral current processes. These include KK modes of the bulk Higgs
as well as doubly charged gauge bosons originating from the \SU symmetry. Since unification
and compactification is to take place at the TeV region we expect $\mu \ra 3e$ to occur
not far from the current experimental limit.  There are many other phenomenologically
interesting signatures in the  production and decay of these particles.
As an example  the  $ U^{-2}$ particle can be produced in a $e^- e^-$ collider as a
resonance.  It then decays into $ l_i l_j$ pair predominantly. The quark decay modes
are absent since they are located on an $SU(2) \times U(1)$ brane. Furthermore
the two $W^-$ channel is forbidden by KK number conservation.
 We can also have  $l^{-} l^{-}\ra W^- V^-$,
where the $V^-$ behaves as an exited $W$. Many such phenomena will be
reserved for future study.

{\bf Acknowledgement.} This work is partially supported by the Natural Science and
Engineering Council of Canada and the National Science
Council of Taiwan, R.O.C (Grant No. 91-2112-M-003-011).
C.H.C. would like to thank Darwin Chang, C.Q. Geng and K. Cheung for insightful discussions.
C.H.C. is grateful for the hospitality and support of
the theory group of TRIUMF, where most of the work was done.


\begin{thebibliography}{99}
\bibitem{notation}
We use the middle Latin alphabets $i,j$ to denote families
$ e_i=e,\, \mu, \, \tau$
and the early alphabets $a,b,c $ to denote electroweak $SU(3)$
indices. In our notation the third component of weak isospin and the hypercharge is
related to the Gell-Mann $\lambda$-matrices  by $T_3=-\frac12\lambda_3,\hs
Y= -\frac{\sqrt3}{2}\lambda_8$ and $Q=T_3+Y$.
\end{thebibliography}
\end{document}